\documentclass[showpacs,twocolumn]{revtex4-1}
\usepackage{amsmath,graphics}
\usepackage[next]{inputenc}
\usepackage[dvips]{epsfig}
\usepackage{hyperref}
\usepackage{color}
\usepackage[dvipsnames]{xcolor}
\usepackage{xcolor}
\usepackage{soul}
\usepackage[normalem]{ulem}
\def\be{\begin{equation}}
\def\ee{\end{equation}}

\def\bi{\begin{itemize}}
\def\ei{\end{itemize}}
\def\bn{\begin{enumerate}}
\def\en{\end{enumerate}}
\def\bea{\begin{eqnarray}}
\def\eea{\end{eqnarray}}
\def\no{\nonumber}
\def\ba{\begin{array}}
\def\ea{\end{array}}
\def\bd{\begin{displaymath}}
\def\ed{\end{displaymath}}

\def\bl{\begin{aligned}}
\def\el{\end{aligned}}

\graphicspath{{Figs/}}

\begin{document}

\title{Real space renormalization of Majorana fermions in quantum nano-wire superconductors}

\author{R. Jafari$^{1,2,3,4,5}$}
\email[]{rohollah.jafari@gmail.com, rouhollah.jafari@physics.gu.se}

\author{A. Langari$^{6,7}$}

\author{ Alireza Akbari$^{1,8,9}$}
\email[]{alireza@apctp.org}

\author{Ki-Seok Kim$^{8,9}$}

\affiliation{$^{1}$Asia Pacific Center for Theoretical Physics (APCTP), Pohang, Gyeongbuk, 790-784, Korea}
\affiliation{$^{2}$School of Physics, Institute for Research in Fundamental Sciences (IPM), Tehran 19395-5531, Iran}
\affiliation{$^{3}$Department of Physics, University of Gothenburg, SE 412 96 Gothenburg, Sweden}
\affiliation{$^{4}$Beijing Computational Science Research Center, Beijing 100094, China}
\affiliation{$^{5}$Department of Physics, Institute for Advanced
Studies in Basic Sciences (IASBS), Zanjan 45137-66731, Iran}
\affiliation{$^{6}$Department of Physics, Sharif University of Technology, Tehran 14588-89694, Iran}
\affiliation{$^{7}$Center of Excellence in Complex Systems and Condensed Matter,
Sharif University of Technology, Tehran 14588-89694, Iran}
\affiliation{$^{8}$Department of Physics, POSTECH, Pohang, Gyeongbuk 790-784, Korea}
\affiliation{$^{9}$Max Planck POSTECH Center for Complex Phase Materials, POSTECH,
Pohang 790-784, Korea}

\begin{abstract}
We develop  the real space quantum renormalization group (QRG) approach for majorana fermions.
As an example we focus  on the Kitaev chain to investigate the topological
quantum phase transition (TQPT) in the one-dimensional 
spinless p-wave superconductor. Studying  the behaviour of local compressibility and
ground-state fidelity, show that the TQPT is signalled by the maximum of local
compressibility at the quantum critical point tuned by the chemical potential.
Moreover, a sudden drop of the ground-state fidelity and the divergence
of fidelity susceptibility at the topological quantum critical point are used as
proper indicators for the TQPT, which signals the appearance of
Majorana fermions. Finally, we present the scaling analysis
of ground-state fidelity near the critical point that manifests the universal
information about the TQPT, which reveals two different scaling behaviors
as we approach the critical point and thermodynamic limit.
\end{abstract}
\date{\today}

\pacs{71.10.Pm, 64.60.ae, 64.70.Tg, 74.90.+n, 03.67.Lx, 74.45.+c}

\maketitle


\section{Introduction \label{introduction}}
Majorana fermions (MFs) 
have attracted intense recent studies in condensed matter
systems\cite{Nayak, Jason}.
Based on exchange statistics, MFs are non-abelian anyons,
in which  particle exchanges  in general do not commute, and they are nontrivial operations\cite{Stern, Chan}.
Furthermore, MFs can be used as qubits in topological quantum computation since they
are intrinsically immune to decoherence \cite{Kitaev20032, Bermudez2}.
Since a MF is its own antiparticle, it must be an equal
superposition of  electron and  hole states. Hence, superconducting systems
are substantial candidates to search for such excitations.
MFs can emerge in systems such as topological insulator-superconductor
interfaces \cite{Fu:2008aa, Linder, Bermudez1, Pawlak}, quantum Hall states with filling factor $5/2$ \cite{Moore},
p-wave superconductors \cite{Read, Nozadze}, and half-metallic ferromagnets
\cite{Duckheim, Chung}.\\

%
\begin{figure*}
\includegraphics[width=0.45\textwidth]{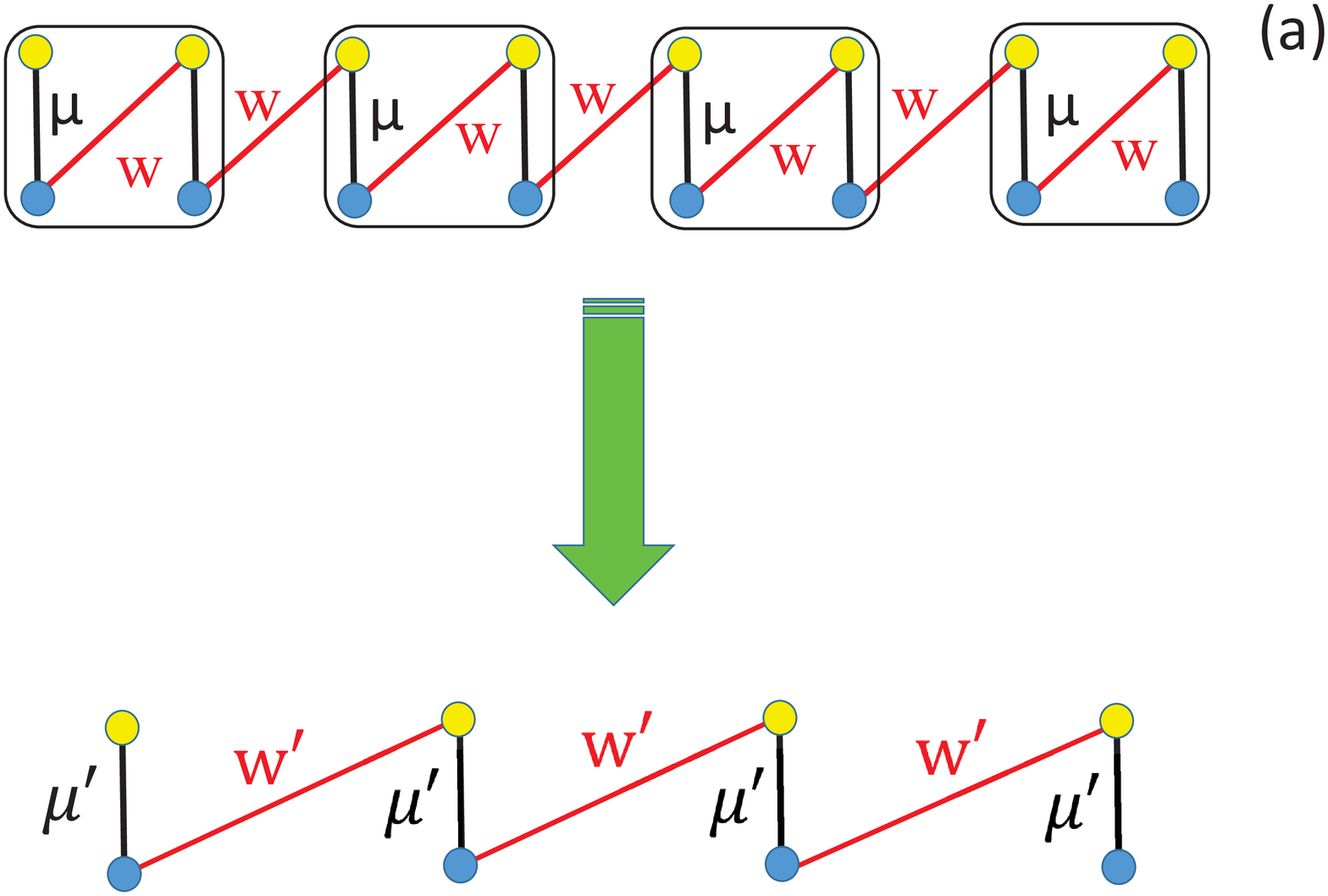}
\hspace{.6cm}
\includegraphics[width=0.45\textwidth]{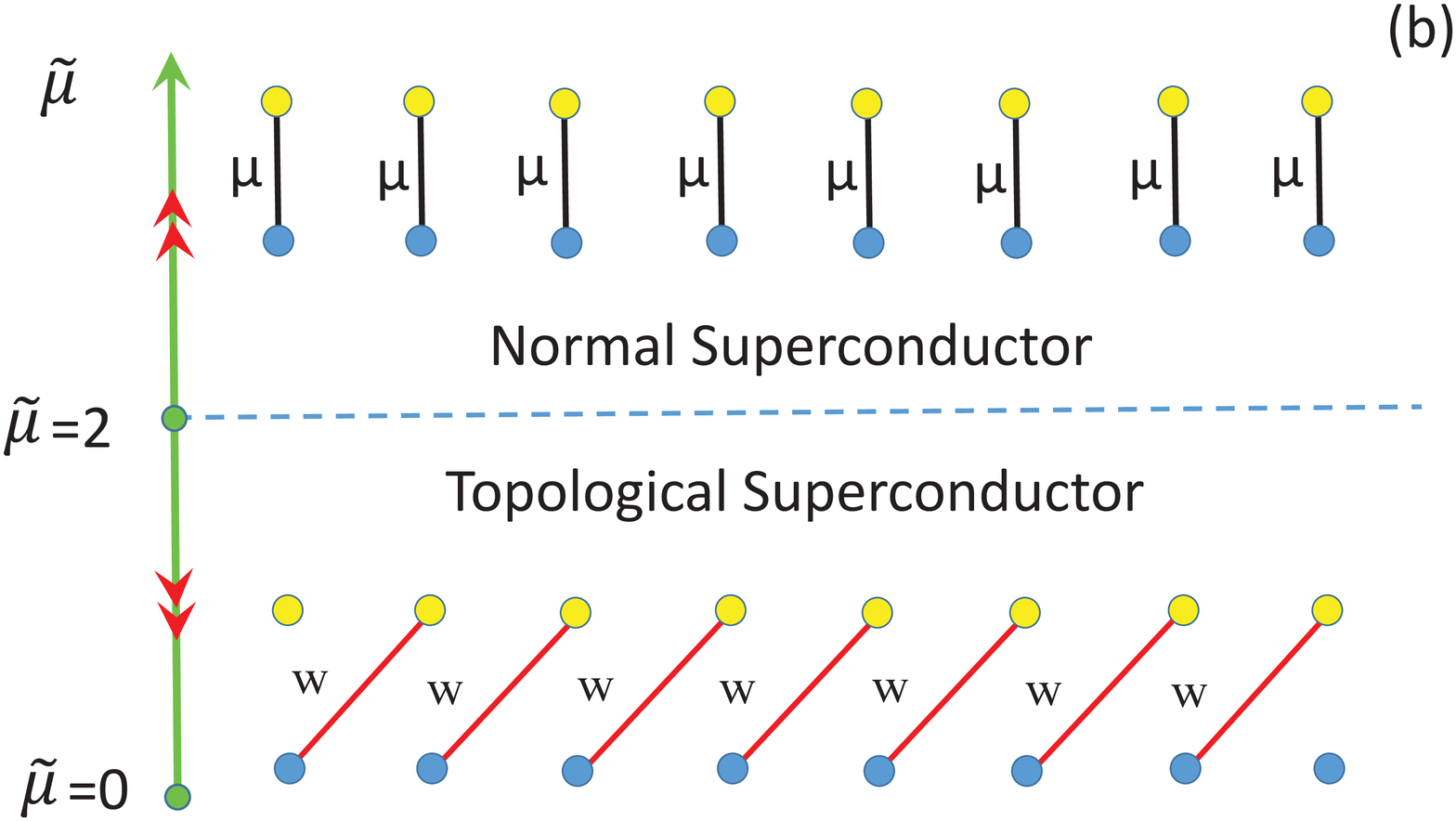}
\caption{(Color online.) (a) Renormalization scheme of Majorana fermions in
nanowire superconductor, (top) where a block of four MFs is mapped to (bottom) a renormalized pair
of MFs. (b) Phase diagram of the one-dimensional Kitaev Model.
In this diagram arrows show the flow of chemical potential under RG iteration.
Each Majorana mode (yellow circle) for a given site
is bound to its partner (blue circle) with strength $\mu$ for the
normal phase,  therefore there are no
unbound modes. However for  the topological phase  there are leaving a free Majorana modes at the ends of a
finite size system.
In the topological superconductor phase the only existing bonds,
with strength $w$, connect each Majorana mode (yellow circle) to its neighbour partner (blue circle).
}
\label{fig1}
\end{figure*}
%

There are various promising proposals for practical realisation of MFs
in one or two dimensional systems.
Among them, the one dimensional (1D) topological
nano-wire superconductors (TSCs) \cite{Lutchyn, Oreg} provide
experimental feasibility for the detection of MFs in hybrid
superconductor-semiconductor wires \cite{Mourik, Deng}.
An egregious feature of a 1D TSC is the edge states (MFs),
which appear at the ends of the superconducting wire.
As shown by Kitaev \cite{Kitaev2001},
MFs can appear at the ends of 1D spinless p-wave superconducting chain when the chemical
potential is less than a finite value, i.e. being in the topological regime.
Recent progress in spin-orbit
coupling research makes it possible to realize Kitaev chain in hybrid systems,
such as superconductor-topological insulator interface \cite{Fu:2008aa},
or semiconductor-superconductor heterostructure
\cite{Sau, Jason, Beenakker}.
In such hybrids, the one dimensional spin-orbit coupling nanowire is proximity
coupled to an ordinary s-wave superconductor \cite{Lutchyn, Oreg}.
It was predicted that there is a topological quantum phase transition (TQPT)
in the system whenever a proper Zeeman field is applied, where the zero-energy modes and MFs appear
in the topological non-trivial phase.
All these experimental observations of the existence of Majorana fermions rely on the fact that the
system is in a topological phase. Moreover, the observation of Majorana fermions has been reported
using scanning tunneling microscopy \cite{Nadj}.

In 1D, the MFs zero energy edge states, can be only found in the chain with open boundary condition, which due to the absence of the translational symmetry an analytical solution is not available. A conventional method to tackle this system is to solve the Bogoliubov-de Gennes equations, diagonalize the Hamiltonian in real space and obtain the energy spectrum as well as the quasi-particle wave functions. However, in the Bogoliubov-de Gennes formalism  by increasing the lattice size, an analytical solution of the wave function and the fidelity are absent.
Most importantly, since the topological phase can not be described within the Landau-Ginzburg symmetry breaking
paradigm,  investigation of such systems is very complicated.
This fact has led to various types of approximations schemes, which can be roughly
classified as variational, perturbative, numerical and renormalization group techniques.
The difficulty of the task suggests  that one should combine various techniques
in order to come as close as possible to the exact solution.
\\
In this paper, we show that how real space quantum renormalization group (QRG) approach is
applicable to MFs in a wire with open ends to acquire the topological phase
transition of the one dimensional p-wave superconductor. We emphasise that this is a technically simple
method, which produces qualitative correct results when properly applied.
Moreover, it is convenient to carry out analytical calculation in the lattice models and they are technically easy to extend to the higher dimensions.
Furthermore, the advantage of the QRG formalism is its capability to evaluate the fidelity and fidelity susceptibility
of a model without referring to the exact ground state of the model.
Particularly, we calculate the local compressibility, ground-state fidelity,
and fidelity susceptibility of the model as the robust geometric probes of quantum criticality.
Ground-state fidelity is a measure, which shows the qualitative change of the ground
state properties without the need to know a prior knowledge of the underlying phases.
The universal scaling properties of fidelity and fidelity susceptibility have been investigated
to extract the universal information of the topological phase transition.
%
\begin{figure}
\centering
\includegraphics[width=0.45\textwidth]{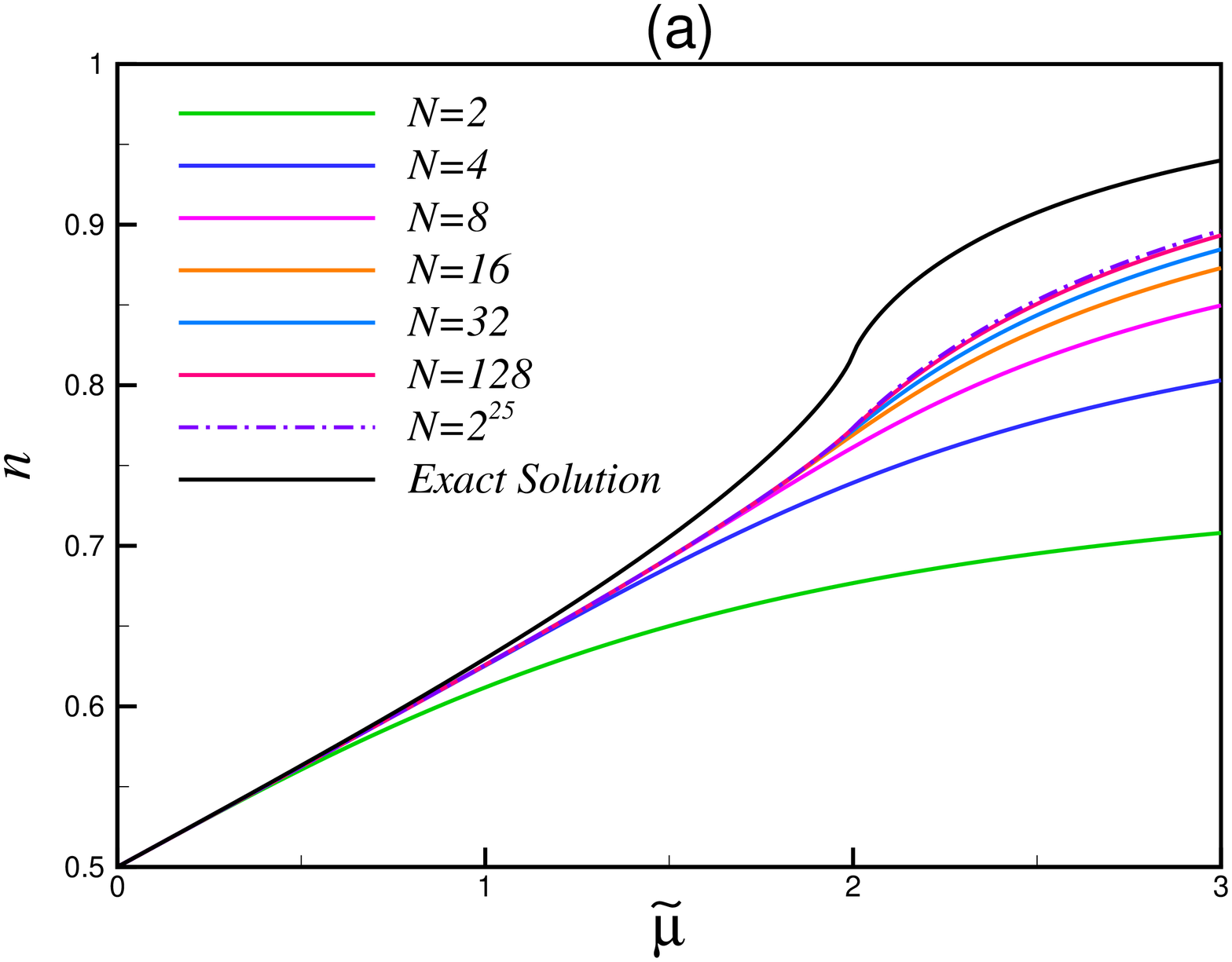}
\\
\includegraphics[width=0.455\textwidth]{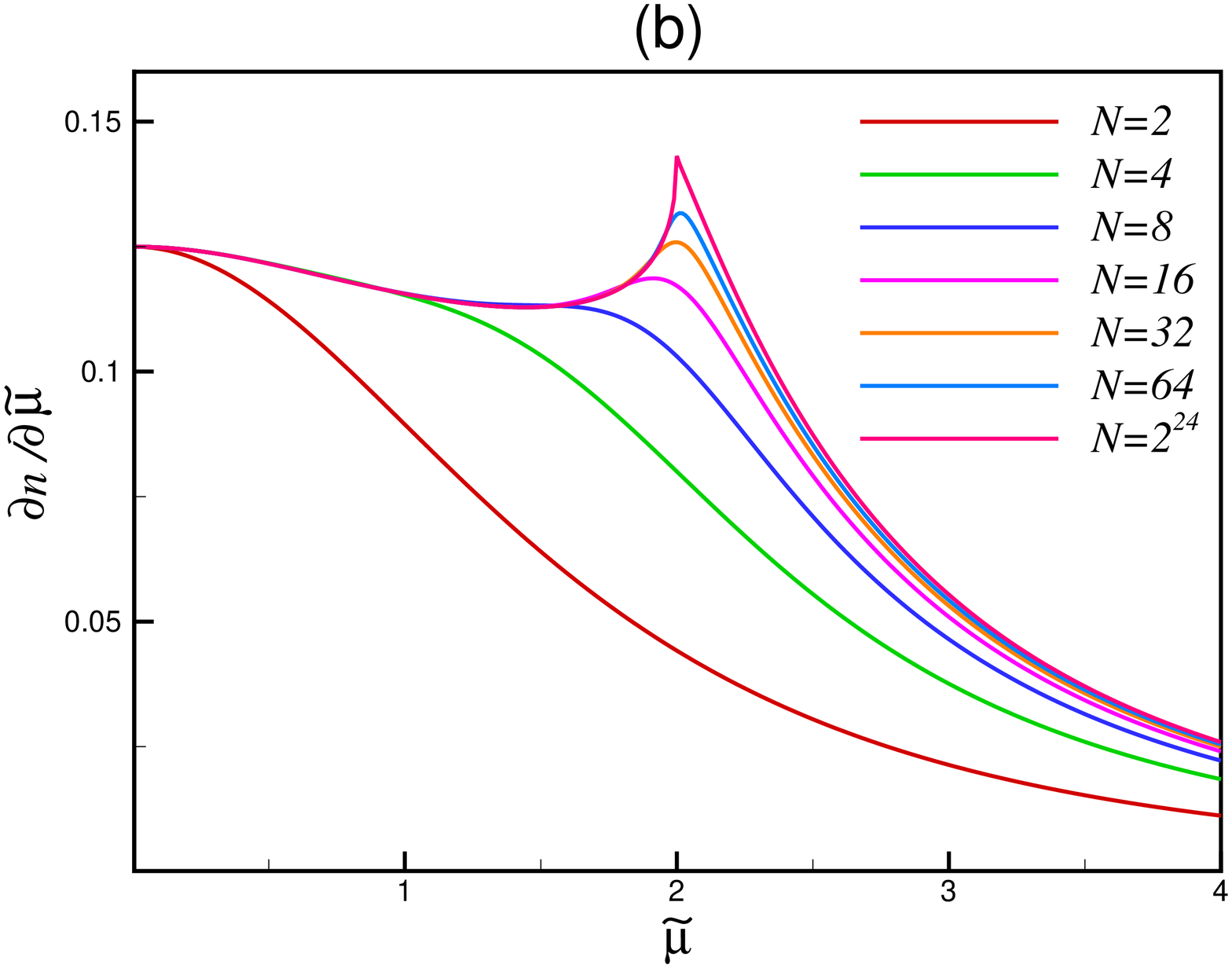}
\caption{(Color online) (a) The particle density $n$ versus chemical potential
$\mu$ for different system sizes (the exact solution data has been taken from
Ref.~\onlinecite{Nozadze}). (b) Derivative of particle density
with respect to chemical potential (compressibility)  versus $\mu$ for different lattice sizes.}
\label{fig2}
\end{figure}
%
The paper is organized as follows: In the next section, the model and majorana fermions
are introduced. In Section \ref{QRG}, the quantum renormalization approach is introduced to
study the ground state phase diagram of the model. In section. \ref{PD}, the density of particle
and compressibility are investigated and section \ref{GF} is dedicated to analysis the ground state
fidelity, fidelity susceptibility, and universal behaviour of the fidelity.
Finally, we will discuss and summarize our results in Section \ref{conclusion}.
\section{One Dimension Quantum Nano-Wire Superconductors \label{model}}
The Kitaev model, which was the first model realizing MFs in a one dimensional lattice \cite{Kitaev2001, Bermudez},
is given
by following Hamiltonian
\be
\bl
\label{eq1}
{\cal H}
=
\sum_{m=1}
\Big[
&
-\mu c^{\dagger}_{m}c_{m}-w(c^{\dagger}_{m}c_{m+1}+c^{\dagger}_{m+1}c_{m})
\\
&
+\Delta(c_{m}c_{m+1}+c^{\dagger}_{m+1}c^{\dagger}_{m})
\Big],
\el
\ee
where $\mu$ is the chemical potential, $c^{\dagger}_{m}$ and $c_{m}$ are the electron creation and annihilation operators on site $m$. The superconducting gap, and hopping integral are defined by $\Delta$ and $w$ respectively.
Since the time-reversal symmetry is broken in Eq.~(\ref{eq1}), we only consider a single value   spin projection, i.e., effectively spinless electrons.\\
By introducing Majorana fermion operators as $a_{n}=c_{n}+c^{\dagger}_{n}$,
and $b_{n}=i(c^{\dagger}_{n}-c_{n})$,
which satisfy the communication relations: $\{a_{m},a_{n}\}=\{b_{m},b_{n}\}=2\delta_{m,n}$ and $\{a_{m},b_{n}\}=0$, the Hamiltonian, Eq.~(\ref{eq1}), takes the following form
\be
\label{eq2}
{\cal H}=\frac{-iw}{2}\sum_{n=1}
\Big[
\tilde{\mu} a_{n}b_{n}+(1-\tilde{\Delta})a_{n}b_{n+1}+ (1+\tilde{\Delta})a_{n+1}b_{n}
\Big],
\ee
where $\tilde{\Delta}=\Delta/w$, and $\tilde{\mu}=\mu/w$.
It is well-known \cite{Kitaev2001} that for the case $|\mu|<2w$, the ground state with MFs is fully realised
and the system is called a topological superconductor, which shows qualitatively different
behavior from the trivial phase, $|\mu|>2w$, without Majorana fermions.
\section{Quantum Renormalization Group \label{QRG}}
QRG  is a method of studying systems with a large
number of strongly correlated degrees of freedom. The main idea of this method
is to decrease or thinning the number of degrees of freedom, so as to retain the
information about essential physical properties of the system and eliminate
those features which are not important for the considered phenomena.
In the real space QRG, which is usually performed on lattice systems with
discrete variables, one can  divide the lattice into blocks which are treated
as the sites of the new lattice.
The Hamiltonian is divided into intra-block and inter-block parts, the former
being exactly diagonalized, and a number of low
lying energy eigenstates are kept to project the full Hamiltonian onto the
new lattice \cite{Delgado, Langari2004, Jafari2006, Jafari20071, Jafari20081, Jafari2010, Jafari2013, Langari1997, Motrunich}.
In the new system there are less degrees of freedom, and the renormalized
couplings are expressed as functions of the initial system's couplings.
Analysing the renormalized couplings (by tracing the flow of coupling constants),
one can determine qualitatively the structure of the phase diagram of the
underlying system, and approximately locate the critical points (unstable fixed points),
and different phases (corresponding to stable fixed points).
To implement the idea of QRG to Majorana fermions in quantum nano-wire
superconductors, the Hamiltonian, Eq.~(\ref{eq2}),
is divided into blocks of four MFs sites, as shown in Fig.~\ref{fig1}(a). In this case,
the total  intra-blocks Hamiltonian is given by
\be
\bl
\label{eq3}
{\cal H_B}
&=\sum_{I=1}^{N/2} h^{B}_{I}
\\
&
=\sum_{I=1}^{N/2}
\frac{-iw}{2}
\Big[
\tilde{\mu}a_{1,I}b_{1,I}+\tilde{\gamma}a_{1,I}b_{2,I}
+\tilde{\lambda}a_{2,I}b_{1,I}
\Big],
\el\ee
where
$h^{B}_{I}$ is  the  sub-Hamiltonian of individual block $I$, with
$
\tilde{\gamma}=1-\tilde{\Delta}
$,
and
$
\tilde{\lambda}=1+\tilde{\Delta}.
$
The remaining part of the Hamiltonian is included in the inter-block part
\be
{\cal H_{BB}}=
\frac{-iw}{2}
\sum_{I=1}^{N/2}
\Big[
\tilde{\mu} a_{2,I}b_{2,I}+\tilde{\gamma}a_{2,I}b_{1,I+1}+\tilde{\lambda}a_{1,I+1}b_{2,I}
\Big],
\ee
where we consider the open boundary chain.
%
\begin{figure}
\centering
\includegraphics[width=0.45\textwidth]{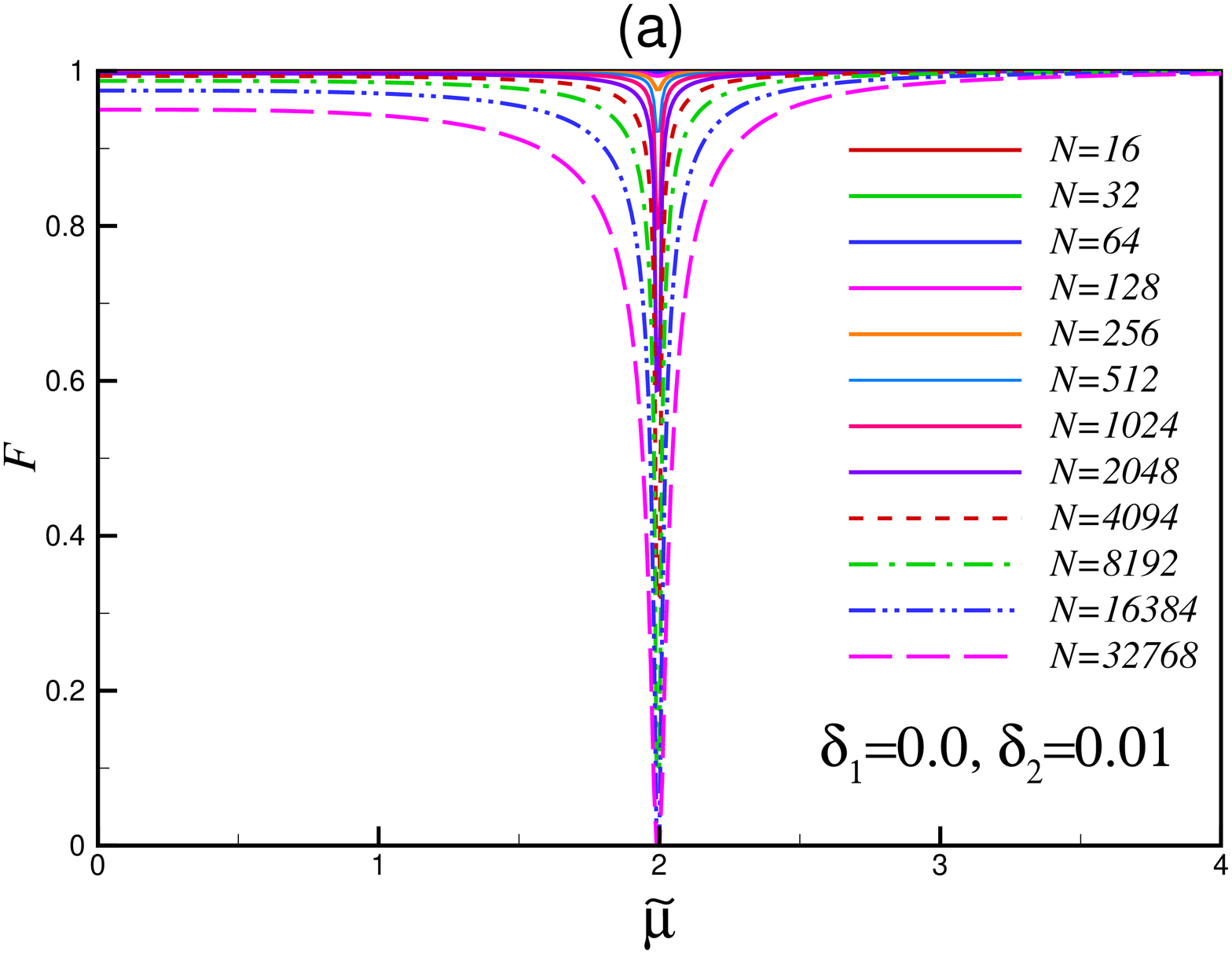}
\\
\includegraphics[width=0.456\textwidth]{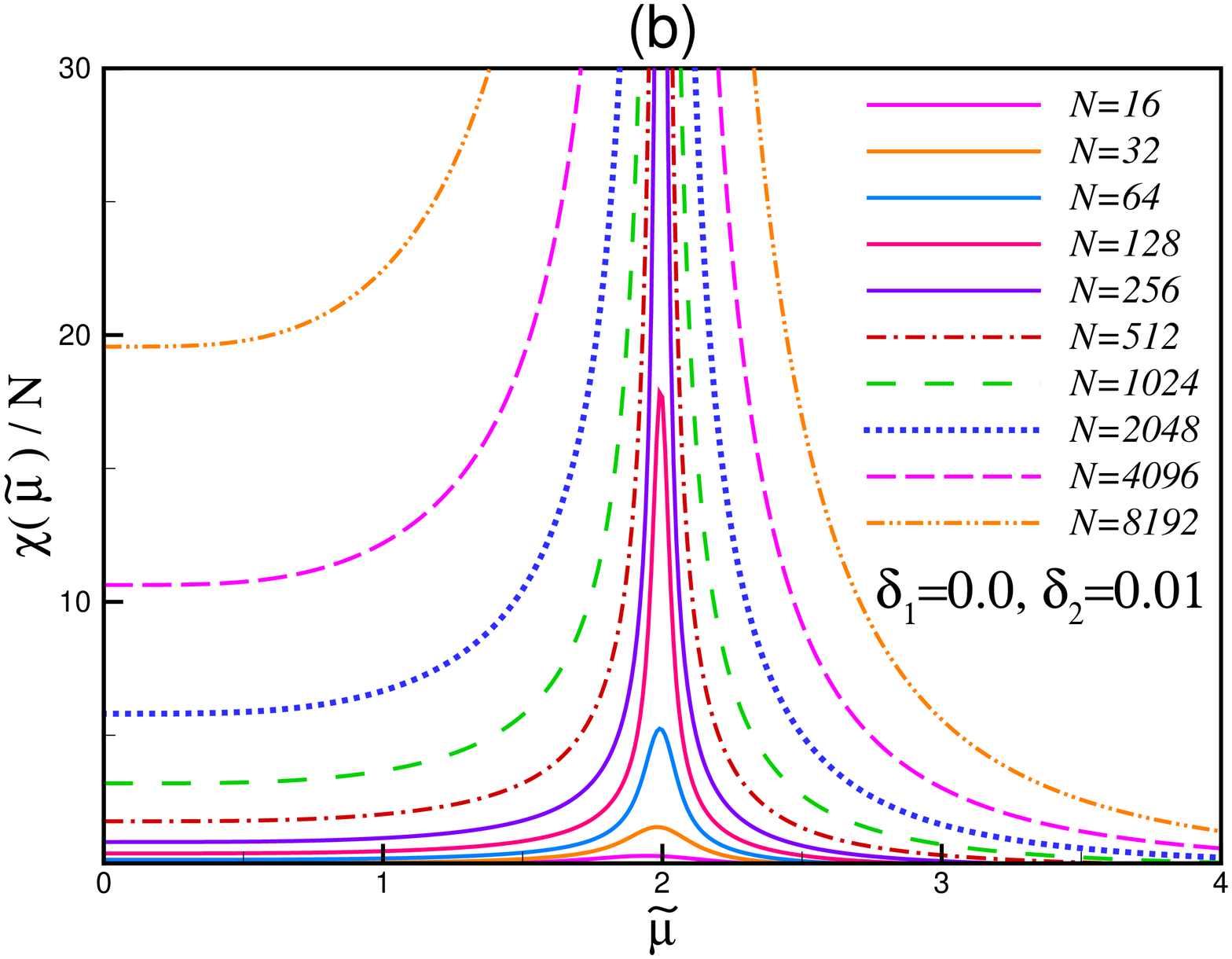}
\caption{(Color online.) (a) Evaluation of the ground state fidelity of
superconducting nanowire under quantum renormalization group for different chain sizes.
 (b) Fidelity
of susceptibility per lattice size for different chain sizes.}
\label{fig3}
\end{figure}
%
The eigenstates and eigenvalues of the block Hamiltonian of four MFs are given by
\be
\bl
\label{eq4}
&
|\psi_{0}\rangle=\frac{\alpha_{0}}{2}(a_{1}-ib_{1})|0^{a,b}_{1},0^{a,b}_{2}\rangle
+\frac{\beta_{0}}{2}(a_{2}-ib_{2})|0^{a,b}_{1},0^{a,b}_{2}\rangle,~
\\
&
|\psi_{1}\rangle=\beta_{1}|0_{1},0_{2}\rangle-
\frac{\alpha_{1}}{4}(a_{1}-ib_{1})(a_{2}-ib_{2})|0^{a,b}_{1},0^{a,b}_{2}\rangle,
\\
&
|\psi_{2}\rangle=\alpha_{1}|0_{1},0_{2}\rangle+
\frac{\beta_{1}}{4}(a_{1}-ib_{1})(a_{2}-ib_{2})|0^{a,b}_{1},0^{a,b}_{2}\rangle,~
\\
&
|\psi_{3}\rangle=\frac{\beta_{0}}{2}(a_{1}-ib_{1})|0^{a,b}_{1},0^{a,b}_{2}\rangle
-\frac{\alpha_{0}}{2}(a_{2}-ib_{2})|0^{a,b}_{1},0^{a,b}_{2}\rangle,
\el
\ee
and
\be
\bl
&
E_{0}=-\frac{w}{2}(\tilde{\mu}+\sqrt{\tilde{\mu}^{2}+4})
;\;\;
E_{1}=-\frac{w}{2}(\tilde{\mu}+\sqrt{\tilde{\mu}^{2}+4\tilde{\Delta}})
;\;\;
\\
&
E_{2}=-\frac{w}{2}(\tilde{\mu}+\sqrt{\tilde{\mu}^{2}-4\tilde{\Delta}^{2}})
;\;\;
E_{3}=-\frac{w}{2}(\tilde{\mu}-\sqrt{\tilde{\mu}^{2}+4}),
\el
\ee
respectively. Here $|0^{a,b}_{1},0^{a,b}_{2}\rangle$ is the vacuum state of MFs in real space and
\be
\bl
\no
&
\alpha_{0}=-\frac{E_{0}/w}{\sqrt{(E_{0}/w)^{2}+1}},~~
\beta_{0}=\frac{1}{\sqrt{(E_{0}/w)^{2}+1}},~~
\\
&
\alpha_{1}=-\frac{E_{1}/w}{\sqrt{(E_{1}/w)^{2}+\tilde\Delta^{2}}},~~
\beta_{1}=\frac{\tilde\Delta}{\sqrt{(E_{1}/w)^{2}+\tilde\Delta^{2}}}.
\el
\ee
The projection operator $P^I_{0}$ for the $I$-th block is defined by
$$
P_{0}^{I}=|\psi_{0}\rangle_{II}\langle\psi_{0}|+|\psi_{1}\rangle_{II}\langle\psi_{1}|,
$$
 and
the renormalization of MFs operators are given by
\be
\bl
&
P_{0}^{I}a_{1,I}P_{0}^{I}=(\alpha_{1}\beta_{0}+\alpha_{0}\beta_{1})a_{I},
\\
&
P_{0}^{I}a_{2,I}P_{0}^{I}=(\beta_{0}\beta_{1}-\alpha_{1}\alpha_{0})a_{I},
\\
&
P_{0}^{I}b_{1,I}P_{0}^{I}=(\alpha_{1}\beta_{0}-\alpha_{0}\beta_{1})b_{I},
\\
&
P_{0}^{I}b_{2,I}P_{0}^{I}=(-\alpha_{1}\alpha_{0}-\beta_{0}\beta_{1})b_{I}.
\el
\ee
It is remarkable that, two different type of Majorana fermion operators in real space
treat nonconformingly under RG transformation. In this respect the effective Hamiltonian is expressed by
\be
{\cal  H}^{eff}=P_{o}({\cal H_{B}}+{\cal H_{BB}})P_{0},
\ee
with total projector
$P_{0}=\bigotimes_{I=1}^{N}P_{0}^{I}$.
Thus, the effective Hamiltonian is obtained
\be
\bl
\label{eq5}
{\cal H}^{eff}=
&
-\frac{i}{2}\sum_{n=1}w^{'}
\Big[\tilde{\mu}^{'} a_{n}b_{n}
+(-1+\tilde{\Delta}^{'})a_{n}b_{n+1}
\\
&
\hspace{2cm}
-(1+\tilde{\Delta}^{'})a_{n+1}b_{n}
\Big]+e_0,
\el
\ee
where $w^{'},~\tilde{\mu}^{'}$, and $\tilde{\Delta}^{'}$ are the renormalized coupling constants defined by the following QRG equations
\be
\bl
\label{eq6}
&
w^{'}=w[\alpha_{0}\beta_{0}
+\tilde\Delta\alpha_{1}\beta_{1}
],
\\
&
\tilde{\Delta}^{'}=-\frac{w}{w^{'}}[\alpha_{1}\beta_{1}
+\tilde\Delta\alpha_{0}\beta_{0}
],
\\
&
\tilde{\mu}^{'}=\frac{(E_{1}-E_{0})-\mu(\alpha^{2}_{1}-\beta^{2}_{0})}{w^{'}},
\el
\ee
and $e_{0}=E_0-\mu\beta_0^2$ is a constant term as a function of coupling constants.
Since the sign of hopping term and superconducting gap are
changed,
the effective Hamiltonian of the renormalized chain
is not exactly similar to the original one. Therefore
to get a self-similar Hamiltonian, we implement the unitary transformation
$a_{n} \longrightarrow (-1)^{n}a_{n}$, and  $ b_{n} \longrightarrow (-1)^{n}b_{n}$,
which is equivalent to $\pi$ rotation of even (or odd) sites around $z$-axis.
We should emphasise  that the commutation relations of MFs operators
do not change under these transformation.\\

The stable and unstable fixed points of the QRG equations are obtained by
solving the following equations
$\tilde{\mu}^{'}=\tilde{\mu}$, and
$ \tilde{\Delta}^{'}=\tilde{\Delta}$.
For simplicity we take $\Delta=w$, for which the Jordan Wigner
transformation reconstructs the model to the Ising model in a transverse field (ITF). For $\tilde{\Delta}=1$ the renormalized couplings reduce to $w^{'}=2w/\sqrt{4+\tilde\mu^{2}}$, and $\tilde{\mu}^{'}=\tilde\mu^{2}/2$.
It is remarkable to mention, although  to compare with  available exact results~\cite{Rams, GU:2010aa, Yu:2009aa} we  look at the QRG for the particular case of $w=\Delta$, our QRG equations 
can be considered for  more general cases  with different values of $w$ and $\Delta$ ($w\neq\Delta$).
The QRG equations show that the stable fixed points are located
at $\tilde\mu=0$ and $\tilde\mu\rightarrow \infty$ while $\tilde\mu_{c}=2$ stands
for the unstable fixed point, which specifies the quantum critical point of the model.
The phase diagram of the model has been shown in Fig.~\ref{fig1}(b).
As depicted, the chemical potential goes to zero for $\tilde\mu<\tilde\mu_{c}$
under QRG transformation, while it scales to infinity in the normal superconductor phase.
The QRG equations, Eq.~(\ref{eq6}), show the flow of $w$ to zero in a normal superconductor,
which represents the renormalization of the energy scale while it goes to $1$ for the
topological superconductor.

Fig.~\ref{fig1}(b) shows that in the topological superconductor phase the only existing
bonds connect $a_{n}$ to its neighbor $b_{n+1}$,  and the Majorana modes at the
ends of the chain are not coupled to anything. It manifests the presence of edge modes
of a topological phase, which is the reason for the ground-state degeneracy that is not
due to a symmetry.
Furthermore,   in the normal superconductor phase, each Majorana mode $a_{n}$ on
a given site is bound to its partner $b_{n}$ with strength $\mu$, leaving no unbound modes,
i.e. no edge state.

The quantum critical exponents associated with this quantum critical point can be obtained from the Jacobian of the QRG transformations by linearizing the QRG flow at the critical point ($\tilde\Delta=1$, and $\tilde\mu=2$),
\be\bl
\label{eq7}
J=
\left(
  \begin{array}{cc}
    \frac{\partial\tilde\mu^{'}}{\partial \mu} & \frac{\partial\tilde\mu^{'}}{\partial \Delta} \\
    \frac{\partial\tilde\Delta^{'}}{\partial \mu} & \frac{\partial\tilde\Delta^{'}}{\partial \Delta} \\
  \end{array}
\right),
\el
\ee
which yields
\be\bl
\label{eq8}
J=
\left(
  \begin{array}{cc}
    2 & -1 \\
    0 & 0 \\
  \end{array}
\right).
\el
\ee
The eigenvalues of the matrix of linearized flow are $\eta_{1}=2$ and $\eta_{2}=0$.
The corresponding eigenvectors in the $|\tilde\mu,\tilde\Delta\rangle$
coordinates are $|\eta_{1}\rangle=|1,0\rangle$, $|\eta_{2}\rangle=|1,2\rangle$. $|\eta_{1}\rangle$
shows the relevant direction which represents the direction of flow of
chemical potential (see Fig.~\ref{fig1}(b)).
We have also calculated the correlation length
exponents at the critical point ${\tilde \mu}_{c}=2$.
In this respect, the correlation length diverges as
$\xi\sim|\tilde{\mu}-2|^{-\nu}$ with the exponent $\nu=1$, which is expressed by
$\nu=\ln[n_{B}]/\ln[{\rm d}\tilde{\mu}^{'}/{\rm d}\tilde{\mu}]|_{\tilde{\mu}_{c}}$,
and $n_{B}=2$ is the number of sites in each block.

%
\begin{figure*}
\centering
\includegraphics[width=0.33\textwidth]{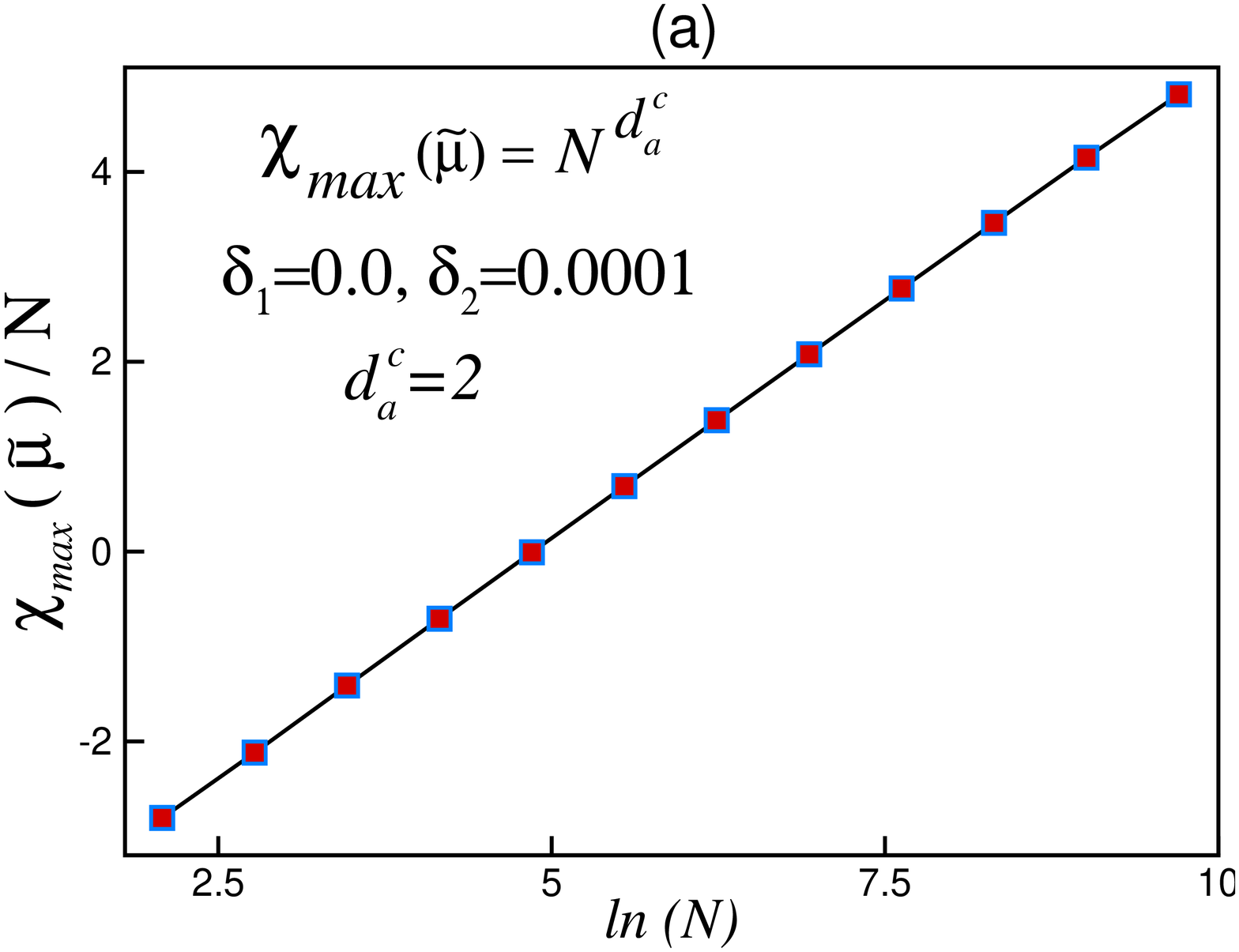}
\includegraphics[width=0.33\textwidth]{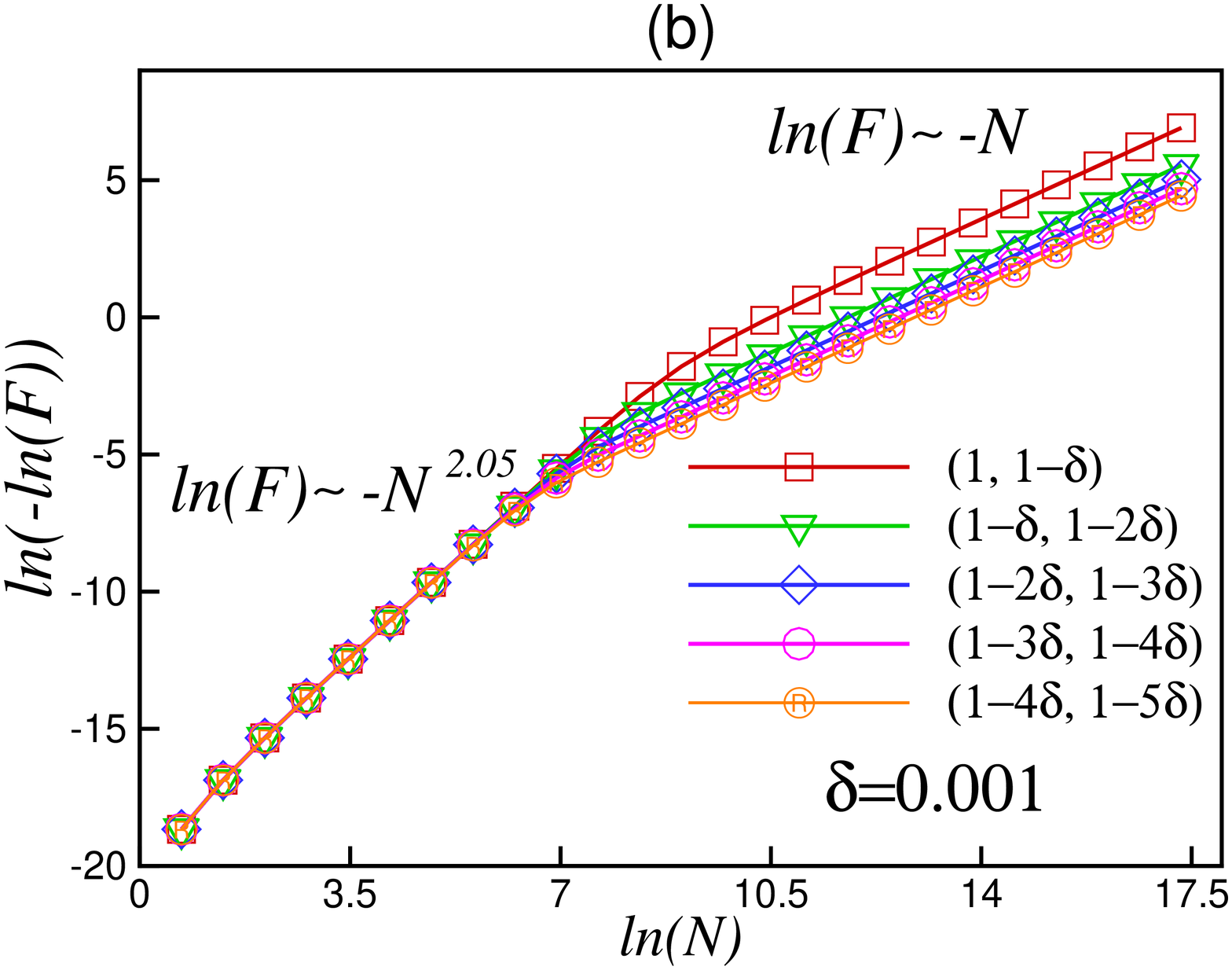}
\includegraphics[width=0.324\textwidth]{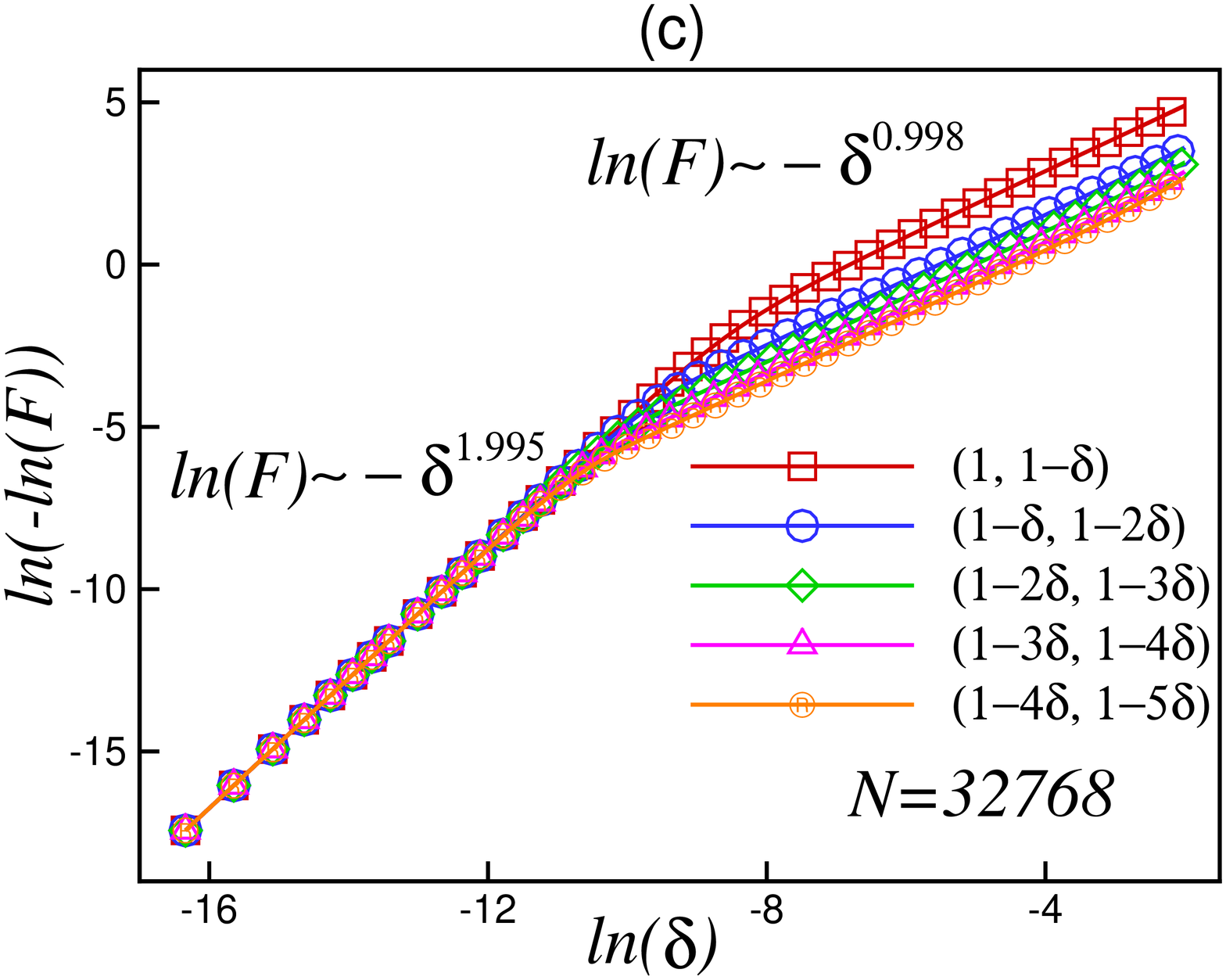}
\caption{(Color online.) (a) The scaling behavior of the maximum
of fidelity susceptibility in terms of system size.
 (b) Scaling behavior
of the ground state fidelity, $\ln(-\ln(F))$, versus $\ln(N)$
close to the critical point $\tilde{\mu}_{c}=2$, for the Kitaev model with fixed
$\delta=0.001$.
 (c)
Scaling behavior of $\ln(-\ln(F))$ versus $\ln(\delta)$ for fixed size
$N=2^{15}$ close to $\tilde{\mu_{c}}=2$.}
\label{fig4}
\end{figure*}
%
\section{particle density \label{PD}}
The average local density of particles on site $n$ is
\be
\bl
\no
n(\tilde\mu)=\frac{1}{N}\sum_{n=1}\langle \Psi_{0}| \frac{(a_{n}-ib_{n})(a_{n}+ib_{n})}{4} |\Psi_{0}\rangle,
\el
\ee
and the ground state of the renormalized chain is related to the ground state of the original one by the
transformation $P_{0}|\Psi_{0}^{'}\rangle=|\Psi_{0}\rangle$.
This leads to the particle density in the renormalized chain
\be
\bl
\label{eq9}
n(\tilde{\mu},N)
&
=\frac{1}{N}\sum_{n=1}\langle O^{'}|P_{0}
\Big[
\frac{(a_{n}-ib_{n})(a_{n}+ib_{n})}{4}
\Big]
P_{0}|O^{'}\rangle
\\
&
=\frac{1}{2}+\gamma(0)n(\tilde{\mu}^{'},\frac{N}{2}),
\el
\ee
where $n(\tilde{\mu}^{'},N/2)$ is the particle density in the
renormalizd chain and $\gamma(0)$ is defined by $(2\alpha^{2}_{1}-1)/2$.

The average of the local density of particles  has been ploted in Fig.~\ref{fig2}(a) for different system sizes. It has been compared with the known exact result \cite{Nozadze}, which shows good agreement
qualitatively. The compressibility,  derivative of the local density of particles with respect
to chemical potential, has been depicted in Fig.~\ref{fig2}(b).
The non-analytic behavior of the particle density at  $\tilde{\mu_{c}}=2$ is high-lighted in the
divergent behavior of the corresponding compressibility. It is to be noted that, although the compressibility has a singularity at the topological critical point in the thermodynamic limit ($N\rightarrow \infty$) this singularity can result from the singularity of density of states \cite{Thouless}. Therefore, to better understand the nature of topological phase transition in the model, the ground state fidelity of the model has been investigated in the next section.
\section{Ground State Fidelity \label{GF}}
In the last few years, fidelity, which is a measure of the distance between
quantum states, has been accepted as a new notion to characterize the
drastic change of the ground states in quantum phase transition  (QPT) point.
Unlike traditional approaches, prior knowledge about the
symmetries and order parameters of the model are not required in
the fidelity notion to find out a  QPT. In addition, fidelity
has an interdisciplinary role, for example, it is related to the density of topological
defects after a quench \cite{Damski2005, Jafari2015a}, decoherence rate of a test
qubit interacting with a non-equilibrium environment
\cite{Damski2011, Jafari2015b}, and orthogonality catastrophe of condensed matter
systems \cite{Anderson1967}. In this section, we implement the formalism
introduced in Refs. [\onlinecite{Langari2012}] and [\onlinecite{Langari2013}]
to calculate the ground state fidelity of the nano-wire superconductor in terms
of QRG.
The advantage of the formalism is its
adroitness to calculate the ground state fidelity of a model without referring to the exact
ground state of the model.

The fidelity of ground state is defined by the overlap between the two ground
state wave functions at different parameter values as follows
\be
\bl
\label{eq10}
F=\langle \Psi_{0}(\tilde\mu-\delta_{1})|\Psi_{0}(\tilde\mu+\delta_{2})\rangle=\langle \Psi_{0}(\tilde\mu_{-})|\Psi_{0}(\tilde\mu_{+})\rangle,
\el
\ee
where $\delta_{j}$ $(j=1,2)$ is a small deviation in the chemical potential.
According to the renormalization transformation $P_{0}|\Psi_{0}^{'}\rangle=|\Psi_{0}\rangle$, the fidelity of renormalized chain are related to the fidelity of original Hamiltonian by $F(\tilde\mu,N)=\gamma_{1}F(\tilde\mu^{'},N/2)$, where
$$
\gamma_{1}=\Big(\alpha_{0}(\tilde\mu_{-})\alpha_{0}(\tilde\mu_{+})+\beta_{0}
(\tilde\mu_{-})\beta_{0}(\tilde\mu_{+})\Big)^{\frac{N}{2}}.
$$
The ground state fidelity of the model has been plotted versus chemical potential in
Fig.~\ref{fig3}(a) for  different system sizes. Obviously, the ground state fidelity
shows a sudden drop at the topological quantum critical point.  Increasing the system size
increases the depth of drop, which manifests unfailing drop in the thermodynamic limit.
To determine more precisely the effect of quantum criticality on fidelity,
we should extract universal information about the transition in addition to providing
the location of the critical point.  The universal information is defined by the critical
exponents and reflects symmetries of the model rather than its
microscopic details. For small system size, the universal information is
explored in the fidelity susceptibility ($\chi$) approach,\cite{Zanardi2006, Rams, Jafari2015a}
which is defined by
\be
\bl
\label{eq11}
\chi=2\lim_{\delta\rightarrow 0}\frac{1-F}{\delta^{2}},
\el
\ee
in which $\delta_1=\delta_2=\delta/2$ is supposed.
Fig.~\ref{fig3}(a) represents fidelity susceptibility (FS) per lattice size ($\chi/N$) versus chemical potential,
for various system sizes.
Although $\chi/N$ does not show divergence for finite lattice sizes, the curves display marked
anomalies with the height of peaks increasing with system size and diverges in the thermodynamics
limit as the critical point is touched. The maximum of FS at finite size lattice $N$ obeys the
scaling $\chi_{max}=N^{d_{a}^{c}}$ where $d_{a}^{c}$ denotes the critical adiabatic dimension\cite{GU:2010aa, Yu:2009aa}.
The scaling analysis of FS is figured out in Fig. {\ref{fig4}(a). It is clearly verified
that the scaling relation $\chi_{max}=N^{d_{a}^{c}}$ is satisfied with $d_{a}^{c}=2$.
It is worth to mention that, the critical adiabatic dimension for the ITF is $d_{a}^{c}=2$ \cite{GU:2010aa, Yu:2009aa}.
Furthermore, it has been shown that in the thermodynamic limit at fixed $\delta$, the fidelity
scaling is given by\cite{Rams}
\be
\bl
\label{eq12}
\ln F(\tilde{\mu},\delta)\sim-N|\delta|^{d\nu}g(\frac{\tilde{\mu}-\tilde{\mu_{c}}}{\delta}),
\el
\ee
where $g$ is a scaling function.
It has been shown that\cite{Rams}  at the critical point,  fidelity is non-analytic in $\delta$  as
$\ln F(\tilde{\mu_{c}},\delta)\sim-N|\delta|^{d\nu}$, while away from critical point for $|\delta|\ll|\tilde{\mu}-\tilde{\mu_{c}}|\ll1$, it behaves as
\be
\bl
\label{eq13}
\ln F(\tilde{\mu}_{c},\delta)\sim-N\delta^{2}|\tilde{\mu}-\tilde{\mu}_{c}|^{d\nu-2}.
\el
\ee
The scaling behavior of fidelity around the critical point, $\tilde{\mu}\approx\tilde{\mu_{c}}$,
is depicted in Fig.~\ref{fig4} (b). As seen, for small system sizes we cover the
well known result \cite{Rams} $\ln F\sim-N^{2}$, reported for ITF model in the finite size scaling
case \cite{Zanardi2006}. However for larger system sizes, we obtain $\ln F\sim-N$ conforming to Eq. (\ref{eq12}). A more detailed analysis shows that the transition between the two regimes takes place when
$N|\delta|\sim1$.
For this purpose in the Fig. \ref{fig4}(c), we  plot the scaling behavior of  $\ln(-\ln F)$ versus $\ln(\delta)$ for the fixed system size. We observe two distinct regimes, namely
for $N|\delta|\ll1$ that  we have $\ln F\sim-\delta^{2}$,
and for $N|\delta|\gg1$ that we find
$\ln F\sim-|\delta|$ in agreement with Eqs.~(\ref{eq12}) and (\ref{eq13}).
It should be mentioned that in our model $d=1$ and scaling verifies the correlation length exponent of $\nu=1$, which exactly corresponds to the correlation length exponent of ITF model.
\section{Summary and conclusions \label{conclusion}}
It is well-known  that the  real space quantum renormalization is applicable to non-exactly
solvable systems such as XXZ model \cite{Delgado,Jafari2006, Jafari20071};
Hubbard model \cite{Bhattacharyya1, Bhattacharyya2, Wang}, as well as,  XY model in a
transverse field \cite{Langari2004}. In this work, we fully formulate quantum renormalization
group for a system of Majorana fermions in the open boundary quantum nano-wire superconductors
to obtain its universal behaviors, such as  phase diagram, compressibility, ground
state fidelity and scaling behavior of fidelity susceptibility.
To compare our analytical approach to exact results, we  concentrate on the special case
with equal  hopping and pairing terms, where the Kitaev model is mapped to the ITF.
We conclude that real space quantum renormalization group
procedure could reproduce exactly the location of the critical point
and the critical exponents. The quantum renormalization group shows that in the topological phase
($\tilde{\mu}<\tilde{\mu}_{c}=2$), the chemical potential goes to zero by renormalization
iteration while in the normal superconductor phase it flows to infinity.
Furthermore, the results layout that the correlation length exponent and
the critical adiabatic dimension are $\nu=1$ and $d^{c}_{a}=2$, respectively. Which corresponds to
their counterpart in the Ising model in transverse field \cite{Rams}.
Finally, we should emphasize that, the quantum renormalization method could be more expedient and more
advantageous than the Bogoliubov-de Gennes formalism to carry out an analytical calculation in the lattice models, specifically in the disorder case and  higher dimensions.

%
\section*{Acknowledgments}
We are grateful to S. Kettemann,  V. Dobrosavljevic, B. Kamble  for fruitful discussions and feedbacks.
The work by  A.A. was supported through  NRF funded by MSIP of Korea (2015R1C1A1A01052411).  A.A. acknowledges support by  Max Planck POSTECH / KOREA Research Initiative (No. 2011-0031558) programs through NRF funded by MSIP of Korea.

\bibliography{References}

\end{document}